\documentclass[%
 reprint,
 amsmath,amssymb,
 prx,
]{revtex4-2}

\usepackage{graphicx}
\usepackage{dcolumn}
\usepackage{bm}
\usepackage{color}
\usepackage{tikz-cd}
\usepackage{amsthm}
\usepackage{amsmath}
\usepackage{mathtools}
\usepackage{enumerate}
\usepackage{amsthm}
\usepackage{xcolor}

\DeclareMathOperator*{\argmax}{argmax}

\newcommand{\diag}{\mathrm{diag}}
\newcommand{\R}{\mathbb{R}}
\newcommand{\dd}{\mathrm{d}}
\newcommand{\dt}{\dd t}

\newcommand{\Img}{\mathrm{Im}}
\newcommand{\Ker}{\mathrm{Ker}}

\newcommand{\Vss}{\mathcal{V}^{ss}}
\newcommand{\D}{\mathcal{E}}

\newcommand{\RnT}{\left( \mathbb{R}^n \right)^*}

\newcommand{\UU}{{U^*}}

\newcommand{\xa}{x^{\textrm{A}}}
\newcommand{\xb}{x^{\textrm{B}}}

\newcommand{\eqnref}[1]{Eq. (\ref{#1})}

\newtheorem{res}{Result}

\makeatletter
\renewcommand*\env@matrix[1][\arraystretch]{%
  \edef\arraystretch{#1}%
  \hskip -\arraycolsep
  \let\@ifnextchar\new@ifnextchar
  \array{*\c@MaxMatrixCols c}}
\makeatother

\begin{document}

\preprint{APS/123-QED}

\title{Riemannian Geometry of Optimal Driving and Thermodynamic Length \\ and its Application to Chemical Reaction Networks}

\author{Dimitri Loutchko}
\email{d.loutchko@edu.k.u-tokyo.ac.jp}
\author{Yuki Sughiyama}
\author{Tetsuya J. Kobayashi}
\homepage{http://research.crmind.net}
\affiliation{Institute of Industrial Science, The University of Tokyo, 4-6-1, Komaba, Meguro-ku, Tokyo 153-8505 Japan}

\date{\today}

\begin{abstract}
It is known that the trajectory of an endoreversibly driven system with minimal dissipation is a geodesic on the equilibrium state space. 
Thereby, the state space is equipped with the Riemannian metric given by the Hessian of the free energy function, known as Fisher information metric.
However, the derivations given until now require both the system and the driving reservoir to be in local equilibrium.
In the present work, we rederive the framework for chemical reaction networks and thereby enhance its scope of applicability to the nonequilibrium situation.
Moreover, because our results are derived without restrictive assumptions, we are able to discuss phenomena that could not been seen previously.
We introduce a suitable weighted Fisher information metric on the space of chemical concentrations and show that it characterizes the dissipation caused by diffusive driving, with arbitrary diffusion rate constants.
This allows us to consider driving far from equilibrium.
As the main result, we show that the isometric embedding of a steady state manifold into the concentration space yields a lower bound for the dissipation when the system is driven along the manifold.
We give an analytic expression for this bound and for the corresponding geodesic, and thereby are able to dissect the contributions from the driving kinetics and from thermodynamics.
Finally, we discuss in detail the application to quasi-thermostatic steady states.
\end{abstract}

\maketitle


\section{Introduction} \label{sec:intro}

In general thermodynamics, the importance of the Hessian of the free energy function as a Riemannian metric has been recognized as early as 1975 by Weinhold \cite{weinhold1975,weinhold1975b,salamon1984} and by Ruppeiner \cite{ruppeiner1979}.
In 1983 it was discovered by Salamon and Berry that this metric controls the dissipation when the system is driven {\it endoreversibly} between equilibrium states \cite{salamon1983}.
For this reason, the concept to determine curves of minimal dissipation from the Riemannian geometry of the parameter space was termed {\it thermodynamic length}.
The derivation required that the driving is given by the explicit form $\dd \eta^i / \dt = k(\eta^i_e - \eta)$ for the extensive variables $\eta^i$ characterizing the equilibrium state of the system and $\eta_e^i$ being the respective variables of the reservoir, with a universal rate constant $k$.
More recently, a derivation using a statistical approach was given by Crooks \cite{crooks2007,crooks2012} based on a cycling driving protocol.
In 2012, Zulkowski, Sivak, Crooks and DeWeese have presented an exact and explicit computation of geodesics for a driven particle diffusing in a one-dimensional harmonic potential \cite{crooks2012b}.

These previous approaches were valid for {\it endoreversible} driving, i.e. for such processes that both the system and the driving reservoir are in local chemical equilibrium and dissipation takes place only at the boundary between the system and the reservoir.
This theory does not directly apply to systems in nonequilibrium or those with algebraic constraints.
Chemical reaction networks (CRN) are an important class of such systems and we extend the concept of thermodynamic  to chemical reaction networks in equilibrium and nonequilibrium steady states.

The importance of chemical reaction networks derives from the fact that they form the basis to model and understand complex chemical processes on a molecular level \cite{mikhailov2017,alon2019}.
They are central to the understanding of biological function as well as in many industrial applications \cite{eungdamrong2004,rabinowitz2012,shinar2010,murray2011,epstein1998,beard2008}, where tight control and high efficiency of the reactions are mandatory.

For CRN, a physically meaningful driving dynamics should be given by $\dd x^i / \dt = k_i(x^i_e - x)$, where $x^i$ is the concentration the $i$-th chemical, $x^i_e$ the concentration of the respective chemical in the reservoir and $k_i$ the diffusion rate constant across the boundary.
This dynamics generalizes the one given by Salamon and Berry and, more importantly, it can drive the CRN away from chemical equilibrium or from a steady state.
Furthermore, models of CRN found in biochemical applications are often open CRN, and as such might not exhibit equilibrium states.
For such networks, the dissipation caused by the driving between steady states is of interest, yet it cannot be determined within the framework available thus far.

The mathematical foundation of CRN theory was laid by Horn, Jackson and Feinberg \cite{horn1972,horn1972b,feinberg1974,feinberg1987,feinberg1988}, and the theory has been refined and developed by employing various mathematical disciplines such as graph theory \cite{okada2016,mochizuki2015,fiedler2015}, homological algebra \cite{hirono2021}, and differential geometry \cite{shinar2009}.
In recent years, it has become clear that CRN are most naturally analyzed using geometrical methods:
Algebraic geometry has led to various important foundational results \cite{craciun2009,gopalkrishnan2014,craciun2019,craciun2020} for CRN with mass action kinetics as well as applications to biologically important networks and phenomena \cite{dickenstein2019,joshi2017,gross2016}.
Meanwhile, information geometry has been successfully used to analyze the geometry of thermodynamics within a stoichiometric compatibility class \cite{yoshimura2021a,yoshimura2021b}.
Finally, both approaches could be merged within the framework of Hessian geometry \cite{shima2007,sughiyama2021,kobayashi2021,sughiyama2022}, generalizing the global structure of the concentration space as studied in algebraic geometry and equipping it with the thermodynamic structure from information geometry as well as including the more classical results \cite{schnakenberg1976,hill1966,hill2005,schmiedl2007,polettini2014,ge2016,rao2016,rao2018} on the thermodynamics of CRN.
In \cite{yoshimura2021a,yoshimura2021b,sughiyama2021,kobayashi2021,sughiyama2022}, the information geometric aspects of Hessian geometry, based on the Bregman divergence, have been thoroughly explored and linked to thermodynamics.
In this work, we focus on the differential geometric aspects of this Hessian geometry.

We show that the dissipation rate caused by the driving of a CRN is determined by the square of the norm $\|j_D \|_{g_X^K}^2$ of the driving flux $j_D$ with respect to the Riemannian metric $\diag (x^1k_1^{-1},\dotsc,x^nk_n^{-1})$ on the concentration space.
This holds arbitrarily far from chemical equilibrium and even for CRN without equilibrium states.
The Riemannian distance $L_X := d_X(x_0,x_1)$ between two concentration vectors $x_0$ and $x_1$ on the concentration space is called {\it thermodynamic length}. Its square is proportional to the minimal dissipation due to the driving in the case that the driving flux $j_D$ and the reaction flux $j_R$ obey the time scale separation $\|j_D\| \gg \|j_R\|$, which allow to ignore the reaction effects.
The geodesic equations can be solved explicitly to yield the optimal trajectory of the system $x(t)$ and the optimal driving protocol $x_e(t)$.
This is {\bf Result \ref{res:1}}, which is derived in Sec. \ref{sec:TDlength}.
The total minimal dissipation $\Sigma_X^{min}$ of this process can be explicitly computed as
\begin{align*}
    \Sigma_X^{min} =  \frac{4}{T} \sum_i k_i^{-1} \left( \sqrt{x^i_1} - \sqrt{x^i_0} \right)^2.
\end{align*}
This is {\bf Result \ref{res:2}} of this work.
Without any assumptions on the time scales of $j_D$ and $j_R$, the minimal total dissipation $\Sigma_{\Vss}^{min}$ caused by the driving of the system along the steady state manifold $\Vss$ between $x_0$ anf $x_1$ on $\Vss$ is proportional to the square length of the Riemannian distance between $x_0$ and $x_1$ on $\Vss$ ({\bf Result \ref{res:3}}).
The optimal system trajectory is thus a geodesic and the optimal driving protocol can be determined from the respective geodesic equations.
The main result ({\bf Result \ref{res:4}}) follows from the isometric embedding of $\Vss$ into concentration space and states that $\Sigma_{\Vss}^{min}$ is bounded from below by $\Sigma_X^{min}$, i.e.
\begin{align*}
    \Sigma_{\Vss}^{min} \geq \frac{4}{T} \sum_i k_i^{-1} \left( \sqrt{x^i_1} - \sqrt{x^i_0} \right)^2.
\end{align*}
This lower bound is universal in the sense that it holds for all possible steady state manifolds irrespective of the reaction kinetics of the CRN.
The {\bf Results \ref{res:1}}-{\bf \ref{res:4}} are derived in Sec. \ref{sec:TDlength}.

In Sec. \ref{sec:Application}, the general results from Sec. \ref{sec:TDlength} are applied to the important class of quasi-thermostatic CRN.
By definition, these are all CRN whose steady state manifold is given by
\begin{align*} 
	\Vss = \{ x \in \R^n_{>0} | \log x - \log x_{ss} \in \Ker[S^T] \},
\end{align*}
where $x_{ss}$ is any steady state and $S$ is the stoichiometric matrix.
This class includes all CRN with equilibrium states that obey the thermodynamics of an ideal solution and all complex balanced nonequilibrium steady states.
An explicit parametrization akin to the exponential family in information geometry is used to derive an explicit expression for the Riemannian metric on the parameter space, which yields numerically solvable geodesic equations for any optimal driving problem.
This parametrization is complemented by a parametrization via conserved quantities, and its physical importance is discussed.
In particular, the parametrization via conserved quantities is an important tool to treat the problem of arbitrary driving protocols with the time scale separation $\|j_R\| \gg \|j_D\|$, which we aim to investigate in the future.
Finally, in Sec. \ref{sec:discussion}, we discuss possible extensions of our approach to systems with non-diagonal Onsager relations.

\subsection*{Setup and notation}

\paragraph*{Notation.} The logarithm and exponential functions of vectors are taken componentwise and the resulting vectors are elements of the linear dual of the original space, i.e.
\begin{align*}
    \exp \left( \sum_i x^ie_i \right) = \sum_i \exp(x^i) e^i,
\end{align*}
where $\{e_i\}$ is a basis for the vector space and $\{e^i\}$ the respective dual basis.
The reason for this convention is that the $\exp$ and $\log$ maps appearing here are Legendre transformations.

By the symbol $\circ$ we denote the Hadamard product between vectors, i.e.
\begin{align*}
    \left( \sum_i x^i e_i \right) \circ \left( \sum_i y^i e_i \right) = \sum_i x^iy^i e_i.
\end{align*}

For the differential geometric formalism, we refer to the textbook \cite{carmo1992}.

\paragraph*{Chemical reaction networks (CRN).}
In this work, we consider a chemical reaction network consisting of $n$ chemicals $X_1,\dotsc,X_n$ and $r$ reactions $R_1,\dotsc,R_r$.
The $j$th reaction is given by
\begin{align*}
    R_j: \sum_{i=1}^n (S_+)^i_j X_i \rightarrow \sum_{i=1}^n (S_-)^i_j X_i
\end{align*}
with nonnegative integer coefficients $(S_+)^i_j$ and $(S_-)^i_j$.
These coefficients determine the reactants and products of the reaction \footnote{Here we suppress the chemicals whose concentrations are kept constant by the coupling to a reservoir, see \cite{sughiyama2021} and \cite{polettini2014} for more details on the role of this coupling.}.
The structure of the network is thus encoded in the $n \times r$ stoichiometric matrix $S = (S^i_j)$ with matrix elements
\begin{align*}
    S^i_j = (S_-)^i_j  - (S_+)^i_j .
\end{align*}
The state of the reaction network is characterized by a vector of nonnegative concentration values $x = (x^1,\dotsc,x^n) \in \mathbb{R}^n_{>0}$, where $x^i$ represents the concentration of the chemical $X_i$.
We denote the concentration space by $X := \mathbb{R}^n_{>0}$.
Finally, the dynamics of the CRN is governed by the equation
\begin{align} \label{eq:dynamics}
    \frac{\dd x}{\dd t} = Sj_R + j_D,
\end{align}
where $j_R=(j_R^1,\dotsc,j_R^r)$ is the vector of reaction fluxes and $j_D = (j_D^1,\dotsc,j_D^n)$ is the vector of the external driving fluxes.
Choosing a kinetic model is tantamount to specifying $j_R$ as a function of $x$.
For any given kinetic model $j_R(x)$, we denote the respective steady state manifold by
\begin{align*}
    \Vss = \{ x \in X | Sj_R(x) = 0 \}.
\end{align*}
The results in this work have general validity as they are not based on the choice of a particular kinetic model for $j_R$.

\paragraph*{Chemical thermodynamics.} 
In chemical thermodynamics, the state of the system is characterized by concentration vectors $x \in X \subset \R^n$ or, equivalently, by chemical potential vectors $\mu$, which live in the dual space $Y:= \RnT$.
We denote  the bilinear pairing between the dual vector spaces by $\langle.,.\rangle$.
The concentration and potential vectors are connected by Legendre duality via the strongly convex free energy function $\varphi(x)$ and its Legendre dual $\varphi^*(\mu)$ as $\varphi^*(\mu) = \max_{x} \left\{ \langle x, \mu\rangle - \varphi(x) \right\}$ and $\mu(x) = \argmax_{\mu} \left\{ \langle x, \mu\rangle - \varphi^*(\mu) \right\}$ and analogous variational characterizations of $\varphi(x)$ and $x(\mu)$.
Additionally, the equality
\begin{align} \label{eq:LegendreDuality}
    \varphi(x) + \varphi^*(\mu) = \langle x, \mu \rangle
\end{align}
is satisfied for the Legendre dual pair of $x$ and $\mu$.
Hereby, the potential $\varphi(x)$ is the Gibbs free energy \footnote{Here, we neglect the constant term due to the coupling to external reservoirs.
See \cite{sughiyama2021} for a thermodynamically thorough discussion of this potential function.}, which takes the form
\begin{align} 
    \label{eq:Gibbs}
    \varphi(x) = \sum_{i=1}^n x^i\left(\mu^0_i + \log x^i -1 \right)
\end{align}
for an ideal dilute solution (or, equivalently, an ideal gas).
The vector $\mu^0 \in \RnT$ is the vector of standard chemical potentials and
we choose the energy scale such that $k_BT =1$.
This explicit form of $\varphi(x)$ will be used in calculations throughout the text.
This yields the $x$-dependence of $\mu$ as
\begin{align} \label{eq:mu_x}
    \mu = \mu^0 + \log x
\end{align}
and, using \eqnref{eq:LegendreDuality}, the explicit form of the Legendre dual potential function \footnote{As for $\varphi(x)$, the constant term arising due to the coupling to external reservoirs is neglected.} as
\begin{align} 
    \label{eq:phi_mu}
    \varphi^*(\mu) = \sum_{i=1}^n \exp( \mu_i - \mu^0_i) = \sum_{i=1}^n x^i.
\end{align}
The strictly convex function $\varphi(x)$ on the concentration space $X$ gives rise to a Riemannian metric via its Hessian
\begin{align*}
    g_X\left(\frac{\partial }{\partial x^i},\frac{\partial }{\partial x^j} \right)  := \frac{\partial^2 \varphi(x)}{\partial x^i \partial x^j}.
\end{align*}
This is known as the Fisher information metric in information geometry \cite{amari2016} and Weinhold or Ruppeiner metric in thermodynamics \cite{weinhold1975,ruppeiner1979}.
Because the spaces we work with have global coordinate systems, the metric can be globally represented by a matrix.
For the specific form of the convex function $\varphi(x)$ given in \eqnref{eq:Gibbs}, the metric $g_X$ is represented by the diagonal matrix
\begin{align} \label{eq:gX}
    g_X = \diag \left(\frac{1}{x^1},\dotsc,\frac{1}{x^n}\right).
\end{align}
To account for the driving kinetics in the next section, we introduce a diagonal weight matrix $K = \diag(k_1,\dotsc,k_n)$ with $k_i \in \R_{>0}$ and the weighted Fisher information metric on $X$ as
\begin{align} \label{eq:gXK}
    g^K_X :=  \diag \left(\frac{1}{k_1x^1},\dotsc,\frac{1}{k_nx^n}\right),
\end{align}
which relates to the Hessian metric $g_X$ as $g^K_X = K^{-1/2} g_X K^{-1/2}$.

The Legendre duality between $X$ and $Y$, together with the Hessian metric $g_X$ and the Bregman divergence is the setup leading up the Hessian geometry of CRN, which was thoroughly explored in \cite{sughiyama2021} and \cite{kobayashi2021}.
Here, we focus on the Riemannian geometry based on the metric $g_X^K$.

\section{Dissipation via thermodynamic length} \label{sec:TDlength}

Throughout this section, we analyze the case of an externally driven CRN between two states $x_0, x_1 \in X$ in finite time $T$.
The full dynamics of the driven CRN is given by \eqnref{eq:dynamics}.
We let $x(t)_{t \in [0,T]}$ denote the resulting integral curve and also have $x_0 = x(0)$ and $x_1 = x(T)$.
The driving is due to the flux $j_D$, which is controlled by the concentrations of chemicals in the reservoir, and the resulting dissipation is evaluated in Sec. \ref{sec:dissipation}. 
If the driving is faster than the chemical reactions, i.e. $\|j_R\| \ll \|j_D\|$, then the dynamics reduces to
\begin{align} \label{eq:dynamics_diffonly}
    \frac{\dd x}{\dt} = j_D
\end{align}
and the optimal trajectory $x(t)_{t \in [0,T]}$ is a geodesic in the concentration space $X$, endowed with the weighted Fisher information metric $g_X^K$.
In this case, the optimal trajectory and the minimal dissipation can be calculated analytically, as shown in Sec. \ref{sec:fastDriving}.

If such a time scale separation does not exist, then the problem of treating an arbitrary driving field $j_D$ is rather intricate as the contributions to the dissipation and to the dynamics generated by $j_D$ and by $j_R$ mix.
Even if the time scale separation $\|j_R\| \gg \|j_D\|$ holds, although this constrains the trajectory $x(t)_{t \in [0,T]}$ to the steady state manifold, the contribution of the dissipation due to chemical reactions cannot be neglected.
Because of the dependence on the details of the particular CRN and of the driving protocol, we do not treat this general case here.

A possibility to circumvent the mixed contributions of $j_D$ and $j_R$ consists in restricting the driving field $j_D$ to the tangent space $T_{x(t)}\Vss$ of the steady state manifold and to impose the initial condition $x_0 \in \Vss$.
This implies that the dynamics is again described by \eqnref{eq:dynamics_diffonly} and that the trajectory $x(t)_{t \in [0,T]}$ is restricted to the steady state manifold.
In this case, as shown in Sec. \ref{sec:driving_onVss}, the optimal trajectory is given by a geodesic on $\Vss$ and can be explicitly computed whenever a parameter manifold for $\Vss$ is known.

\subsection{Dissipation} \label{sec:dissipation}

We assume that the system is in contact with an external reservoir and can exchange all its chemicals $X_1,\dotsc,X_n$ via diffusion with the reservoir, with diffusion rate constants $k_1,\dotsc,k_n$.
The reservoir is thereby characterized by the time-dependent concentration vector $x_e(t) = (x_e^1(t),\dotsc,x_e^n(t))$ and the corresponding chemical potential vector $\mu^e(t)$, which satisfy $\mu^e(t) = \mu^0 + \log x_e(t)$.
The diffusion flux is given by Fick's law, i.e. the flux vector field on $X$ is 
\begin{align} \label{eq:jD_explicit}
    j_D(x,t) =  K \left(x_e(t) - x(t) \right),
\end{align}
where $K = \diag(k_1,\dotsc,k_n)$ is the matrix of diffusion rate coefficients.
In the following, we suppress the time-dependence of the variables for the sake of clarity of exposition.
In addition, we assume that the external reservoir potential vector is close to the system potential vector, i.e. $\mu^e \approx \mu$.
Then the relation
\begin{align} \label{eq:driving_potential}
    \mu^e - \mu &= \frac{\partial \mu}{\partial x} \left(x^e - x \right) = g_X \left(x^e - x \right)
\end{align}
holds \footnote{This can be directly seen from the Taylor expansion $x_e^i - x^i = \exp(\mu^e_i - \mu^0_i) - \exp(\mu_i - \mu^0_i) = \exp(\mu_i - \mu^0_i) \left( \exp(\mu^e_i - \mu_i) -1 \right) = x^i \left( \mu^e_i - \mu_i + \mathcal{O}\left((\mu^e_i - \mu_i)^2 \right) \right)$}, where $g_X$ is the Fisher information metric introduced in \eqnref{eq:gX}.
This leads, using Eqs. (\ref{eq:jD_explicit}) and (\ref{eq:driving_potential}), to the dissipation rate at the boundary as
\begin{align} 
    \nonumber
    \sigma_D &= \langle j_D, \mu^e - \mu \rangle  =  \langle j_D, g_X  \left(x^e(t) - x \right) \rangle \\
    \nonumber
    &=  \langle j_D, K^{-\frac{1}{2}} g_X  K^{-\frac{1}{2}} j_D \rangle. \\
    \label{eq:sigma_D}
    &=  \| j_D \|_{g^K_X}^2,
\end{align}
where $g_X^K$ is the weighted Fisher information metric introduced in \eqnref{eq:gXK}.
This expression is the reason for introducing the weights to the standard Fisher information metric:
The dissipation rate at the boundary between the system and the reservoir is given by the squared norm of the driving vector $j_D(x,t)$.
Together with \eqnref{eq:dynamics_diffonly}, this allows to relate the dissipation rate to the squared speed of the integral curve $x(t)$ and the total dissipation to its energy (here, {\it energy} is used in the context of differential geometry, which has no relation to the thermodynamical energy).

\subsection{Optimal fast driving} \label{sec:fastDriving}

If the time scale separation $\|j_R\| \ll \|j_D\|$ holds, which we call the {\it fast driving} regime, the optimal system trajectory $x(t)$, the corresponding driving protocol $x_e(t)$ and the minimal dissipation can be determined analytically.
In the fast driving regime, the dynamics of the driven system obeys \eqnref{eq:dynamics_diffonly} and therefore the dissipation rate is given by \eqnref{eq:sigma_D}:
\begin{align*}
    \sigma_D = \left\| \frac{\dd x}{\dt} \right\|^2_{g_X^K}
\end{align*}
The total dissipation $\Sigma_X$ is therefore
\begin{align} \label{eq:sigma_tot_X}
    \Sigma_X = \int_0^T \left\| \frac{\dd x}{\dt} \right\|^2_{g_X^K} \dt.
\end{align}  
Applying the Cauchy-Schwarz inequality to the integral yields
\begin{align} \label{eq:CauchySchwarz}
    \Sigma_X \geq \frac{1}{T} \left( \int_0^T \left\| \frac{\dd x(t)}{\dt} \right\|_{g_X^K} \dt \right)^2 = \frac{1}{T}L_X^2,
\end{align}
where $L_X$ is the length of the curve $x(t)$.
For the trivial weight matrix $K=I$, this is known as {\it thermodynamic length} \cite{salamon1983} and we also use this term for the lengths of curves on spaces endowed with the weighted Fisher metric.
The equality in \eqnref{eq:CauchySchwarz} holds if and only if the speed of the curve, $\left\| \frac{\dd x}{\dt} \right\|_{g_X^K}$, is constant.
In this case, $x(t)$ is a geodesic.
If it is a minimal length geodesic, then both the length $L_X$ of the curve and its energy $\Sigma_X$ are minimized.
In this case, $L_X$ is equal to the distance between $x_0$ and $x_1$ on the space $X$, i.e.
\begin{align*}
    L_X^{min} = d_X(x_0,x_1) = \min_{\substack{\gamma : [0,T] \rightarrow X \\ \gamma(0) = x_0 \\ \gamma(T) = x_1}} \int_0^T  \left\| \frac{\dd \gamma(t)}{\dt} \right\|_{g_X^K} \dt.
\end{align*}
For the driving field given by diffusion, i.e. $j_D = K(x_e -x)$, and the weighted Fisher information metric $g_X^K$, the geodesic equations are
\begin{align*}
    \frac{\dd^2 x^i}{\dd t^2} -\frac{1}{2x^i} \left( \frac{\dd x^i}{\dd t} \right)^2 = 0
\end{align*}
for $i=1,\dotsc,n$, cf. Appendix \ref{app:geodesics}.
They have the explicit solution
\begin{multline} \label{eq:opt_driving}
    x^i(t) = - \left( \sqrt{x^i_1} - \sqrt{x^i_0} \right)^2 \frac{t}{T} \left( 1- \frac{t}{T} \right) + \\
     + x^i_1 \frac{t}{T} + x^i_0 \left( 1 - \frac{t}{T} \right)
\end{multline}
for $t \in [0,T]$.
The second term of the expression describes a linear trajectory in concentration space, which would yield the shortest path in the standard Euclidean metric, and the first term shows the correction due to the non-Euclidean geometry of the concentration space.
Interestingly, the optimal curve $x(t)$ does not depend on the details of the driving kinetics, i.e. it is identical for all possible values of the diffusion rate constants $k_i$.
In other words, the optimal curve is determined purely by the thermodynamics of the system via the Hessian $g_X$.
This is the first main result of this section:
\begin{res} \label{res:1}
The optimally driven curve $x(t)$ in the fast driving limit is given by \eqnref{eq:opt_driving}.
It does not depend on the diffusion rate constants but only on the thermodynamics of the system.
The optimal driving protocol $x_e(t)$ is given by
\begin{align} \label{eq:opt_protocol}
    x_e(t) = K^{-1} \frac{\dd x}{\dt} + x(t),
\end{align}
which follows from Eqs. (\ref{eq:dynamics_diffonly}) and (\ref{eq:jD_explicit}).
\end{res}
With the explicit solution at hand, \eqnref{eq:sigma_D} yields the dissipation rate and \eqnref{eq:sigma_tot_X} the total dissipation in the driving process and thus the second main result:
\begin{res} \label{res:2}
The optimally driven system under fast driving has the constant dissipation rate
\begin{align*}
    \sigma_D^{min} = \frac{4}{T^2} \sum_i k_i^{-1} \left( \sqrt{x^i_1} - \sqrt{x^i_0} \right)^2
\end{align*}
and the minimal total dissipation is
\begin{align} \label{eq:Sigma_min_X}
    \Sigma_X^{min} =  \frac{4}{T} \sum_i k_i^{-1} \left( \sqrt{x^i_1} - \sqrt{x^i_0} \right)^2.
\end{align}
\end{res}
The results are not only significant in the fast driving regime but they provide lower bounds for dissipation when the trajectory $x(t)$ is restricted to the steady state manifold, as is shown in the following section.

\subsection{Driving between arbitrary steady states} \label{sec:driving_onVss}

In this section we consider the case that the driving field $j_D$ is constrained to the tangent space $T_{x(t)} \Vss$ of the steady state manifold with the initial condition $x_0 \in \Vss$.
This implies that the integral curve $x(t)$ is restricted to $\Vss$ and therefore the term $Sj_R(x(t))$ in \eqnref{eq:dynamics} vanishes for all points on the curve by definition of $\Vss$.
Thus, the dynamics is given by
\begin{align*}
    \frac{\dd x}{\dt} = j_D 
\end{align*}
and the dissipation rate is given by $\sigma_D = \| \frac{\dd x}{\dt} \|_{g_X^K}$ as before.
Now, the integral curve is restricted to $\Vss$ and by the same argument as before, the minimal length curve on $\Vss$ between $x_0$ and $x_1$ will minimize the dissipation.
This can be made precise (and accessible to direct computation as demonstrated in Sec. \ref{sec:Application}) as follows:
Without loss of generality we assume that $\Vss$ is connected and thus parametrized by a connected manifold $M$, i.e. there is an embedding
\begin{align*}
    f: M \rightarrow X
\end{align*}
with $\Img[f] = \Vss$.
Defining the metric on $M$ as the pullback
\begin{align*}
    g_M^K := f^*g^K_X
\end{align*}
makes the embedding $f$ isometric and thus the dissipation rate can be written as
\begin{align*}
    \sigma_D = \left\| \frac{\dd x}{\dt} \right\|^2_{g_X^K} = \left\| \frac{\dd m}{\dt} \right\|^2_{g_M^K},
\end{align*}
where $f(m) = x$.
The total entropy production is bounded from below by
\begin{align*}
    \Sigma_{\Vss} = \int_0^T \left\| \frac{\dd m}{\dt} \right\|^2_{g_M^K} \dt \geq \frac{1}{T} L_{\Vss}^2,
\end{align*}
where $L_{\Vss}$ is the length of the curve $m(t)$ on $M$, which is bounded from below by the distance $d_{\Vss}\left(m_0,m_1\right)$ between $m_0 := f^{-1}(x_0)$ and $m_1 := f^{-1}(x_1)$ on $M$, i.e.
\begin{align*}
    \nonumber
    L_{\Vss} \geq d_{\Vss}\left(m_0,m_1\right) = \min_{\substack{\gamma : [0,T] \rightarrow M \\ \gamma(0) = m_0 \\ \gamma(T) = m_1}} \int_0^T  \left\| \frac{\dd \gamma(t)}{\dt} \right\|_{g_{\eta}^K} \dt.
\end{align*}
This leads to the following result:
\begin{res} \label{res:3}
    In the case that the system is driven on the steady state manifold, the dissipation $\Sigma_{\Vss}$, which is caused by the driving, is minimized if and only if $m(t)$ is a geodesic on $M$.
    In this case, the minimal dissipation reaches its lower bound, which is given by
    \begin{align*}
        \Sigma_{\Vss}^{min} = \frac{1}{T} d_{\Vss}\left(m_0,m_1\right)^2.
    \end{align*}
    The integral curve on the concentration space is $x(t) = f(m(t))$ and the corresponding driving protocol is given explicitly by
    \begin{align*}
         x_e(t) &= K^{-1} J_f \frac{\dd m}{\dt} + f(m),
    \end{align*}
    which follows from \eqnref{eq:opt_protocol} and where $J_f$ is the Jacobian of $f$ at $m(t)$.
    This result holds independently of any time scale separation between $j_D$ and $j_R$.
\end{res}

\begin{figure} [ht]
    \centering
    \includegraphics[width=0.5\textwidth, clip]{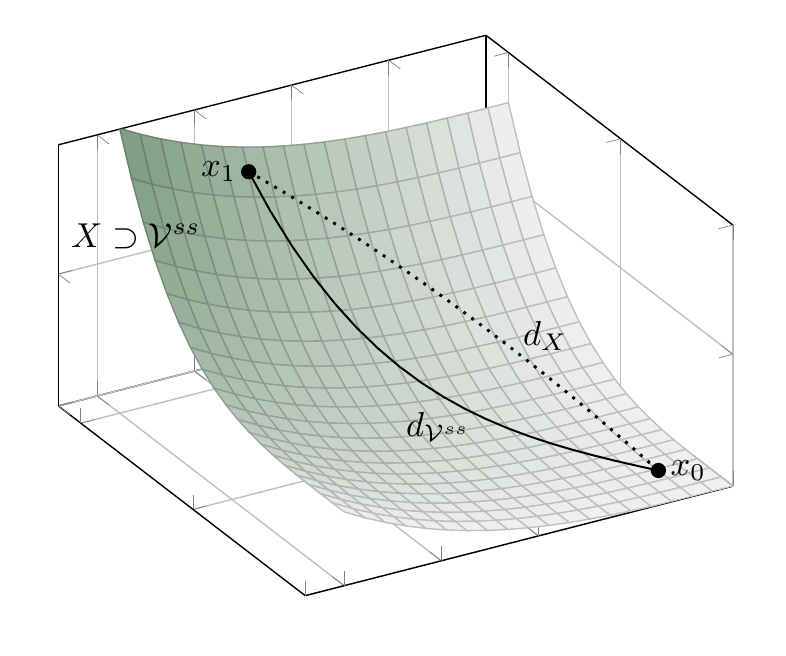}
    \caption{Illustration of the approach to the lower bound for the total dissipation $\Sigma_{\Vss}^{min}$.
    The isometric embedding of $\Vss$ (represented by the green surface) into the concentration space $X$ allows to compare the lengths of the geodesics $d_{\Vss}$ and $d_X$ on $\Vss$ and $X$ between the two points $x_0,x_1 \in \Vss$ as $d_{\Vss} \geq d_X$.
    The relations $\Sigma_X^{min} = d_X^2/T$ and $\Sigma_{\Vss}^{min} = d_{\Vss}^2/T$ show that $\Sigma_{\Vss}^{min} \geq \Sigma_X^{min}$.
    The latter can be explicitly computed, cf. \eqnref{eq:Sigma_min_X}, and the former is a lower bound for the dissipation when the system is driven on $\Vss$ with any protocol, thus yielding Result \ref{res:4}.
    }
    \label{fig:driving_Vss}
\end{figure}

\noindent Finally, the isometry of the embedding $f$
implies that 
\begin{align*}
    d_X (x_0,x_1) \leq d_{\Vss}\left(m_0,m_1\right)
\end{align*}
and thus $\Sigma_X^{min} \leq \Sigma_{\Vss}^{min}$.
This is illustrated in Fig. \ref{fig:driving_Vss}. 
We are led to the main result of this work:
\begin{res} \label{res:4}
    The total dissipation $\Sigma_{\Vss}$ for a driving process on the steady state manifold is bounded from below by
    \begin{align*}
        \Sigma_{\Vss} \geq \frac{4}{T} \sum_i k_i^{-1} \left( \sqrt{x^i_1} - \sqrt{x^i_0} \right)^2.
    \end{align*}
    This lower bound is achieved if and only of the system is driven on the optimal curve for the fast driving regime, given by \eqnref{eq:opt_driving}.
\end{res}
Note that nonequilibrium steady states have a nonzero dissipation rate caused by continuously ongoing chemical reactions.
Our approach does not take this dissipation into account, but only gives a recipe for minimizing the dissipation caused through the driving.

For an arbitrary CRN, the parametrization of the steady state manifold $f:M \rightarrow X$ is, in general, not available.
In fact, even when $j_R$ is given by mass action kinetics, the determination of steady states involves highly nontrivial algebraic geometry and must be carried out for each CRN individually \cite{dickenstein2016}.
Yet, the Result \ref{res:2} holds irrespectively of any of the complicated details of $\Vss$.

For the class of quasi-thermostatic CRN, an explicit parametrization of $\Vss$ is available and the Result \ref{res:3} can be made precise and accessible for direct calculation.
This is presented in the next section.

\section{Driving in quasi-thermostatic CRN} \label{sec:Application}

Quasi-thermostatic CRN, as defined by Horn and Jackson in \cite{horn1972}, are characterized by the following particular form of the steady state manifold
\begin{align} \label{eq:quasiTS_Vss1}
	\Vss = \{ x \in X | \log x - \log x_{ss} \in \Ker[S^T] \},
\end{align}
where $x_{ss}$ is a particular solution of the equation $Sj_R(x_{ss}) = 0$.
The quasi-thermostatic property derives its importance from the fact that all equilibrium states of ideal solutions and complex balanced steady states of CRN with mass action kinetics are quasi-thermostatic.
The class of quasi-thermostatic CRN is, however, even broader because there exist quasi-thermostatic CRN that belong to neither of the two classes mentioned above \cite{horn1973}.
In the following, we give two possible parametrizations of quasi-thermostatic steady states and illustrate each with an explicit calculation of the geodesic and minimal dissipation based on Result \ref{res:3}.

\subsection{Exponential Parametrization of Quasi-thermostatic CRN} \label{sec:toric}

By choosing a basis $u_1^*,\dotsc,u_q^*$ of $\Ker[S^T]$ and writing $\UU = (u^*_1,\dotsc,u^*_q)$ for the $n \times q$ matrix of basis vectors, the characterization in \eqnref{eq:quasiTS_Vss1} can be rewritten as
\begin{align} \label{eq:quasiTS_Vss}
	\Vss = \{ x_{ss} \circ \exp(\UU \eta^*)| \eta^* \in \R^q \}.
\end{align}
Thus, we obtain the parametrization of $\Vss$ via the embedding
\begin{align} \label{eq:toric_param}
	f: \R^q &\rightarrow X \\
	\nonumber
	\eta^* &\mapsto x_{ss} \circ \exp(\UU \eta^*).
\end{align}
This structure has been thoroughly studied in algebraic geometry \cite{craciun2009,craciun2020}, where $\Vss$ was shown to be a toric variety and, using the matrix-tree theorem explicit formulae for $x_{ss}$ were obtained.
In the particular case that $\Vss$ is the equilibrium manifold of a closed CRN, $x_{ss}$ is given by $\exp (-\mu^0)$.
In this work, the analytification $\Vss$ of the toric variety from algebraic geometry is used.
We call $\Vss$ a toric manifold.

\begin{figure}[ht]
\includegraphics[width=0.5\textwidth, clip]{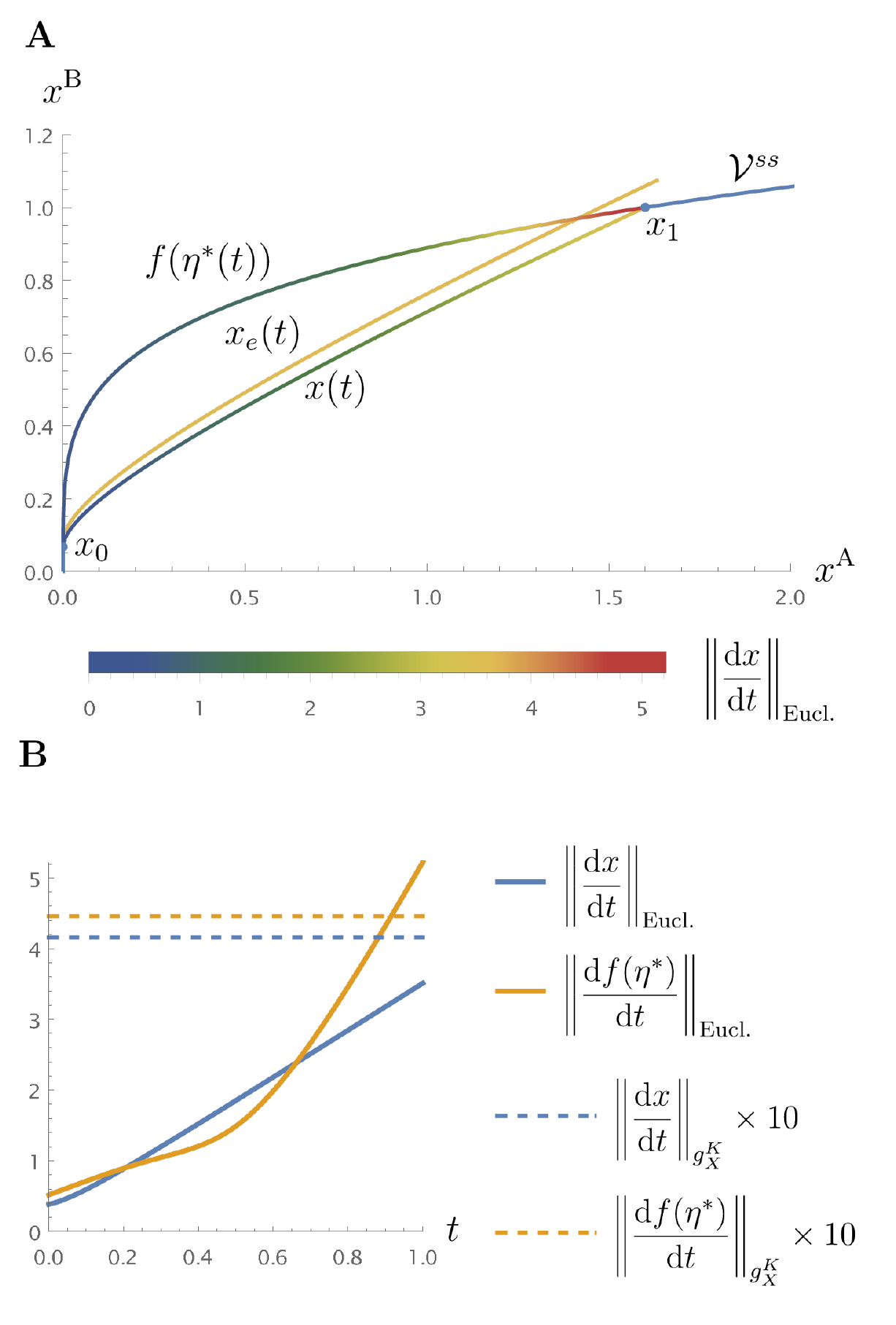}
\caption{
Illustration of optimal driving of the $\textrm{A} \rightleftharpoons 4\textrm{B}$ CRN, using the numerical values $\xa_{ss}=0.1$, $\xb_{ss}=0.5$, $k_{\textrm{A}}=100$, $k_{\textrm{B}}=20$, $\eta^*_0 = -2$, $\eta^*_1 = \log 2$ and $T=1$; with $x_0 = f(\eta^*_0)$ and $x_1 = f(\eta^*_1)$.
{\bf A}. Trajectories of the system being driven from $x_0$ to $x_1$.
The lower curve $x(t)$ is the optimal curve on the whole concentration space, given by \eqnref{eq:opt_driving} and the curve $x_e(t)$ (yellow) is the corresponding optimal driving protocol, given by \eqnref{eq:opt_protocol}. 
The curve $f(\eta^*(t))$ is the optimal curve on the steady state manifold $\Vss$ (light blue).
The curves $x(t)$ and $f(\eta^*(t))$ are color coded according to the speed of the driving, given by the Euclidean norm of the tangent vector $\frac{\dd x}{\dt}$.
In {\bf B}, the Euclidean norms, which give the speed of the driving, and the $\|.\|_{g_X^K}$ norms (their 10-fold values are plotted here for better visibility), which govern the dissipation, of the tangent vectors $x(t)$ and $f(\eta^*(t))$ are plotted for the course of the driving.
}
\label{fig:example}
\end{figure}

We now treat the Riemannian geometry of the quasi-thermostatic steady state manifold $\Vss$.
The pullback of the metric $g_X$ to $\R^q$ can be computed as
\begin{align*}
    g_{\R^q}^K &= f^* g_X^K = \left( \frac{\partial x}{\partial \eta^*} \right)^T g_X^K \left( \frac{\partial x}{\partial \eta^*} \right) \\
        &= J_f^T g_X^K J_f.
\end{align*}
With the Jacobian explicitly given by $J_f = D \UU$, where $D := \diag (x^1,\dotsc,x^n)$.
Note that $D$ can be explicitly expressed in the $\eta^*$ coordinates by using \eqnref{eq:toric_param}.
We obtain the desired metric on the parameter space $\R^q$ as
\begin{align} \label{eq:metric_RqT}
	g_{\R^q}^K = (\UU)^T K^{-1/2} D K^{-1/2} (\UU)^T.
\end{align}
Although we could not solve the geodesic equations on $\R^q$ in an analytically closed form, they can be solved numerically for any given CRN.
The following example illustrates this.
\paragraph*{Example.}
Consider the following nonlinear chemical reaction network
\begin{equation*} \label{dia:CRN1}
\begin{tikzcd}[nodes in empty cells]
    \textrm{A} \ar[r, leftharpoonup, shift left=.35ex] & 4 \textrm{B} \ar[l, leftharpoonup, shift left=.35ex],
\end{tikzcd}
\end{equation*}
with stoichiometric matrix $S = (-1~4)^T$.
The space $\Ker[S^T]$ is spanned by the vector $u^* = (4~1)$, the space of $\eta^*$ variables is one-dimensional, i.e. it is given by $\R^1$, and the metric $g_{\R^1}^K$ is given by
\begin{align*}
	g_{\R^1}^K = 16k_{\textrm{A}}^{-1}\xa_{ss} \exp (4 \eta^*) + k_{\textrm{B}}^{-1} \xb_{ss} \exp (\eta^*)
\end{align*}
according to \eqnref{eq:metric_RqT}.
The geodesic equation for $\eta^*$ is 
\begin{align} \label{eq:geodesic_eta}
	\frac{\dd^2 \eta^*}{\dt^2} + \left( \frac{64k \exp(3 \eta^*) + 1}{32 k \exp(3\eta^*) +2} \right) \left( \frac{\dd \eta^*}{\dt} \right)^2 = 0,
\end{align}
with $k:= \frac{k_B \xa_{ss}}{k_A \xb_{ss}}$.
Fig. \ref{fig:example}A shows the geodesic on the steady state manifold determined by \eqnref{eq:geodesic_eta} and the geodesic on the space $X$, which is determined by \eqnref{eq:opt_driving}, with the numerical parameter values given in the figure caption.
The curves are color coded by the speed of the driving, which is given by the Euclidean norms of $\frac{\dd x}{\dt}$ and $\frac{\dd f(\eta^*)}{\dt}$, respectively.
The speed of the curves is also shown in Fig. \ref{fig:example}B and as a function of time.
The speed increases with increasing norm of the coordinate $x$ for both curves, whereby the system driven on $\Vss$ has the greater acceleration.
As both curves are geodesics, their speed with respect to the metric $g_X^K$ must be constant, as is confirmed by the numerical results shown in Fig. \ref{fig:example}(B).
The total dissipation for the optimal driving along $\Vss$ is $\Sigma^{min}_{\Vss} = T \left\| \frac{\dd f(\eta^*)}{\dt} \right\|_{g^K_X}^2 \approx 0.20$, with the lower bound $\Sigma_X^{min} = T \left\| \frac{\dd x}{\dt} \right\|_{g^K_X}^2 \approx 0.17$.

For linear CRN, i.e. CRN where all the reaction complexes consist of exactly one chemical, the geodesic on $X$, given by \eqnref{eq:opt_driving}, and the geodesic on $\Vss$ coincide (see Appendix \ref{app:crooks}).
In this case the equality $\Sigma^{min}_{\Vss} = \Sigma_X^{min}$ holds and $\Sigma^{min}_{\Vss}$ can be computed by \eqnref{eq:Sigma_min_X}.
In order to illustrate the difference between $\Sigma^{min}_{\Vss}$ and $\Sigma_X^{min}$ we have thus chosen an example with a strong nonlinearity.
But even in the nonlinear case, $\Sigma_X^{min}$ provides a good estimate for $\Sigma^{min}_{\Vss}$.
This is not only the case for the given example but for various other CRN that we have analyzed numerically.
In future work, we aim to analytically quantify the error in estimating $\Sigma^{min}_{\Vss}$ by $\Sigma_X^{min}$ by relating it to the curvature of $\Vss$.

\subsection{Parametrization via conserved quantities}

\begin{figure}[ht]
\centering
\includegraphics[width=0.5\textwidth, clip]{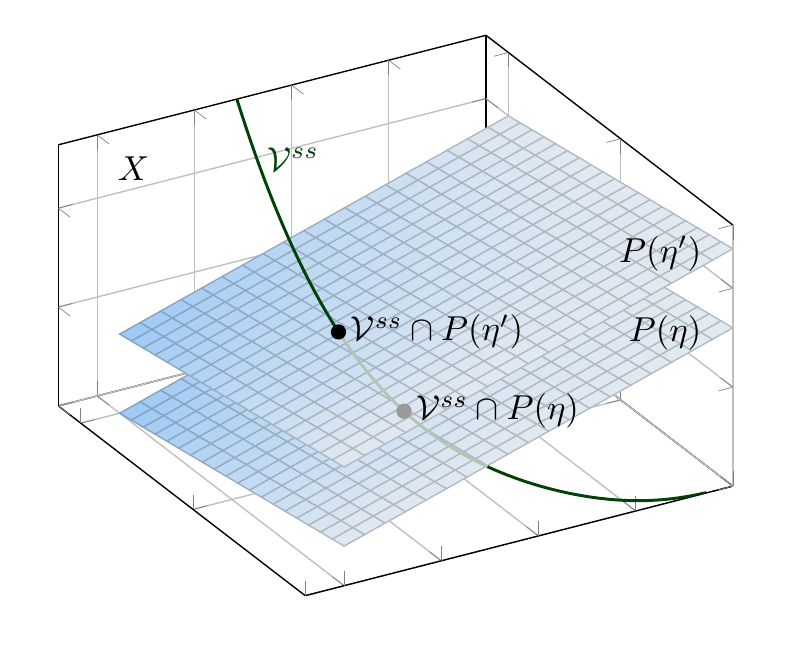}
\caption{Illustration of the parametrization of points on a manifold of quasi-thermostatic steady states $\Vss$ via the vector of conserved quantities $\eta$.
The codimension of $\Vss$ in $X$ is the dimension of the stiochiometric polytope $P(\eta)$ and therefore a zero dimensional intersection set is expected in general.
The dual orthogonality from Hessian geometry \cite{sughiyama2021,kobayashi2021} or, alternatively, Birch's theorem from algebraic geometry \cite{craciun2009}, ensure that the intersection consists of exactly one point for each $\eta \in \D$.
As such, this structure is a folitation of $X$, with the base manifold being $\Vss$ and the leaves being $P(\eta)$.
In other words, $\Vss$ is the moduli space of stiochiometric compatibility classes of the CRN.
In the figure, it is illustrated how the $P(\eta)$ and $P(\eta')$ are parallel affine spaces in $X$, that cover all of $X$ and are indexed by points on $\Vss$.}
\label{fig:dual_param}
\end{figure}

In addition to the parametrization of the steady state manifold $\Vss$ of quasi-thermostatic CRN as a toric manifold, given by \eqnref{eq:toric_param}, the manifold $\Vss$ can be parametrized by the vector of conserved quantities $\eta = (\UU)^T x$, with the matrix $\UU$ introduced in the previous section and $x \in \Vss$ \footnote{Because $\UU$ spans $\Ker[S^T]$ and thus $\frac{\dd (\UU)^T x}{\dt} = (\UU)^T S j_R = 0$, the quantities $(\UU)^T x$ are conserved by the reaction dynamics of the CRN.}.
This is based on Birch's theorem \cite{pachter2005, craciun2009}, which states that the intersection of $\Vss$ with the stoichiometric polytope $P(\eta) := \left\{ x \in X | (\UU)^T x = \eta \right\}$ exists and is unique, i.e. the map $(\UU)^T: \Vss \rightarrow \R^q$, defined by $x \mapsto (\UU)^T x$, has a unique inverse given by
\begin{align} \label{eq:cons_param}
	h: \D &\rightarrow X \\
	\nonumber
	\eta &\mapsto \Vss \cap P(\eta),
\end{align}
where $\D := (\UU)^T \Vss$ is the image of $\Vss$.
This is the parametrization of quasi-thermostatic steady states by the vector of conserved quantities.
The geometrical reason for the uniqueness of this intersection is the generalized orthogonality between $\Vss$ and $P(\eta)$, as discussed in \cite{kobayashi2021,sughiyama2021}.
Fig. \ref{fig:dual_param} illustrates this.

Although the map $h$ cannot be given in analytically closed form, its Jacobian can be computed.
First note that the map $h^{-1} f$ is the coordinate change from $\eta^*$ to $\eta$ variables and its Jacobian is given by
\begin{align*}
    J_{h^{-1}f} = \left( \frac{\partial \eta^*}{\partial \eta} \right) = (\UU)^T D \UU,
\end{align*}
where $f$ is the map given in \eqnref{eq:toric_param}.
Using $J_f = D\UU$ leads to
\begin{align} \label{eq:Jh}
    J_h = J_f J_{f^{-1} h} = J_f J_{h^{-1} f}^{-1} = D\UU \left( (\UU)^T D \UU \right)^{-1}.
\end{align}
Thus, defining the weighted Fisher information metric $g_{\D}^K$ on $\D$ by
\begin{multline} \label{eq:g_eta_explicit}
    g_{\D}^K = \left( \frac{\partial \eta^*}{\partial \eta} \right)^T g_{\R^q}^K \left( \frac{\partial \eta^*}{\partial \eta} \right) = J_{f^{-1} h}^T g_{\R^q}^K J_{f^{-1} h} \\
    = \left( (\UU)^T D \UU \right)^{-1} (\UU)^TK^{-1}D\UU \left( (\UU)^T D \UU \right)^{-1}
\end{multline}
makes the parametrization $h:\D \rightarrow \Vss$ isometric \footnote{Note that an analytical expression for $g_{\D}^K$ in the $\eta$ coordinates is not available in general, because the matrix $D$ requires an analytically closed form of the map $h$.}.
This is summarized in the following diagram
\begin{equation} \label{dia:isometries_Veq}
    \begin{tikzcd}[nodes in empty cells]
    \left(\D , g_{\D}^K \right) \ar[r, shift left=.55ex, "h"] &
    \left(\Vss, \left.g^K_X\right|_{\Vss} \right) \ar[d, hook, "\iota"] \ar[r, shift right=.55ex, swap, "f^{-1}"] \ar[l, shift left=.55ex, "(\UU)^T"]
    &  \left( \mathbb{R}^q , g_{\R^q}^K \right)  \ar[l,shift right=.55ex, swap, "f"] \\
     & \left(X, g^K_X \right). \ar[ul, shift left=.55ex, bend left=25, "(\UU)^T"] & 
\end{tikzcd}
\end{equation}
With the respective metrics $g^K_{\D}$, $g^K_{\R^q}$ and $\left.g^K_X\right|_{\Vss}$, the maps in the upper row of the diagram are isometries and thus the dissipation through finite time driving can be computed on either of the three spaces.
The embedding $\iota: \Vss \rightarrow X$ is isometric and allows for the explicit computation of lower bounds for  the total dissipation due to the driving as stated in Result \ref{res:4}.

Finally, we remark that the coordinates $\eta$ and $\eta^*$ are Legendre dual, see \cite{kobayashi2021} for details.

\subsection{Applicability of the two parametrizations}

The metric $g^K_{\R^q}$ does have an analytically closed expression in the $\eta^*$-coordinates via Eqs. (\ref{eq:toric_param}) and (\ref{eq:metric_RqT}).
Therefore, the space $\R^q$ of $\eta^*$-parameters, equipped with the metric $g^K_{\R^q}$, is suitable for explicit computations.
The example in Sec. \ref{sec:toric} can be generalized to arbitrary quasi-thermostatic CRN, i.e. the geodesic equations in $\eta^*$-coordinates on $\R^q$ can be explicitly written and numerically solved for a given set of parameters to obtain the optimal system trajectory, the optimal driving protocol and the minimal total dissipation. 

\begin{figure}[ht]
\includegraphics[width=0.5\textwidth, clip]{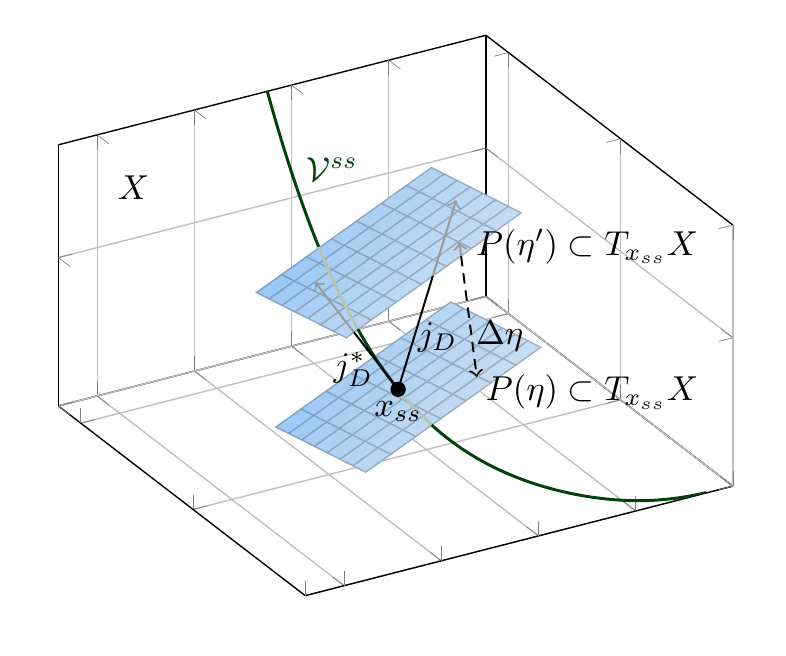}
\caption{The construction of a driving vector field $j^*_D$ tangent to $\Vss$ from an arbitrary driving field $j_D$ such that both fields generate the same slow dynamics on $\Vss$.
At each point $x_{ss} \in \Vss$, the vector field $j_D$ will drive the system away from $\Vss$ in general.
However, if the time scale separation $\|j_R \| \gg \|j_D \|$ holds, the steady state reached after such a perturbation will be determined by the change in the vector of conserved quantities $\Delta \eta = \eta' - \eta$, i.e. by the pushforward $J_{(\UU)^T} j_D$ to the tangent space $T_{(\UU)^Tx_{ss}} \D$.
All tangent vectors at $x_{ss}$, which lie in $P(\eta')$ have the same pushforward.
Among them, there is exactly one, which is tangent to $\Vss$.
This is the desired driving field $j_D^*$.
Note that in this construction, both stoichiometric polytopes $P(\eta)$ and $P(\eta')$ lie in $T_{x_{ss}}X$ and not in $X$.
}
\label{fig:j_D}
\end{figure}

In contrast, there is no analytically closed expression for the metric $g_{\D}^K$ on the space $\D$ of $\eta$-coordinates.
However, the vector of conserved quantities $\eta$ has an intuitive physical meaning and is a natural tool to analyze the case of the time scale separation $\|j_R \| \gg \|j_D \|$.
In the case of arbitrary driving fields $j_D$, the time scale separation $\|j_R \| \gg \|j_D \|$ results in a slow time scale dynamics on $\Vss$, which is determined by the pushforward $(\UU)^T x(t) = \eta(t)$ of the trajectory $x(t)$ to the space of conserved quantities.
While the detailed analysis of this scenario is dependent on the details of the CRN, the slow dynamics on $\Vss$ can be reproduced by the trajectory $h((\UU)^T x(t))$.
The corresponding driving protocol is given by
\begin{align} \label{eq:xe_slow}
    x_e^{slow}(t) = K^{-1} J_h (\UU)^T \frac{\dd x(t)}{\dt} + h((\UU)^T  x(t)).
\end{align}
However, this formula requires the evaluation of the function $h$.
This can be circumvented by transforming the original driving field $j_D$ into a driving field $j_D^*$ on the tangent space $T\Vss$ of the steady state manifold.
For each $x \in \Vss$ and $t \in [0,T]$, the change in steady state, which is the state that the system will eventually relax to after the disturbance caused by $j_D(x,t)$, is described by the change of the vector of conserved quantities.
The latter quantity is given by the pushforward $J_{(\UU)^T} j_D(x,t)$ to the tangent space $T_{(\UU)^Tx}\D$.
There is one and only one vector field $j_D^*(x,t)$ on $T_x\Vss$, which causes the same change in the vector of conserved quantities as $j_D(x,t)$.
It must be given by the pushforward of $J_{(\UU)^T} j_D(x,t)$ to $\Vss$, i.e.
\begin{align*}
    j_D^*(x,t) := J_h J_{(\UU)^T} j_D (x,t).
\end{align*}
Using the explicit formula for the Jacobian $J_h$ given in \eqnref{eq:Jh}, one obtains the desired vector field as
\begin{align*}
    j_D^*(x,t) =  D\UU \left( (\UU)^TD\UU \right)^{-1} (\UU)^T j_D(x,t).
\end{align*}
This approach is illustrated in Fig. \ref{fig:j_D}.
One verifies by a direct calculation that $J_{(\UU)^T} j^*_D = J_{(\UU)^T} j_D$ and therefore $j_D$ and $j_D^*$ indeed induce the same slow dynamics on $\Vss$.
Moreover, in the case that $j_D$ is already a vector field on the tangent space of $\Vss$, the equality $j^*_D = j_D$ holds.
Therefore, when considering arbitrary driving fields and the time scale separation $\|j_R \| \gg \|j_D \|$, this construction plays an important role.
In future work, it has to be supplemented by an analysis of the dissipation caused by the relaxation of the system to $\Vss$ caused by $j_R$. 

\section{Discussion and Outlook} \label{sec:discussion}

In this work, we have analyzed the application of the Riemannian geometry with a weighted Fisher information metric to the dissipation in driven chemical reaction networks.
This geometry should be thought of as an infinitesimal and weighted version of the Hessian geometry established in \cite{sughiyama2021,kobayashi2021}, where information geometric aspects based on the Bregman divergence were analyzed.
Inspired by the classical results on thermodynamic length, we have shown that the dissipation rate is given by the speed of the integral curve if either the diffusion is fast or the driving generates curves that lie on the steady state manifold.
Thereby, we have presented a mathematically and physically precise derivation of this formula based on the physical model of diffusion of chemicals between the system and the reservoir.
Moreover, we were able to explicitly include the diffusion rate constants into the calculation.
This was the reason to introduce weights in the various Fisher information metrics.

The careful mathematical setup allowed us, almost effortlessly, to obtain several new results for the dissipation in driven chemical reaction networks.
If the diffusion of the driving is faster than the chemical reactions, a complete analytical solution was given:
The optimal curve in concentration space, the optimal driving protocol and the minimal total dissipation were determined.
Interestingly, it was found that the optimal curve does not depend on the diffusion rate constants, i.e. it is determined purely by the thermodynamics of the system, whereas the optimal driving protocol depends on the kinetics.

Then, without any assumption on the time scales, if the driving proceeds along the steady state manifold, the geodesic equations were formulated on the respective parameter space.
For the case of quasi-thermostatic steady states, the equations were derived explicitly and thus they can be numerically solved for any given CRN.
More importantly, the isometric embedding of the steady state manifold into the concentration space was used to obtain a lower bound for the total dissipation.
This bound is universal as it is independent of the particular kinetic model $j_R(x)$ and is valid for any steady state manifold.
In all numerical examples we investigated thus far, this bound turned out to provide a good estimate for the actual total dissipation.
Thereby the estimation error seems to grow with the curvature of the steady state manifold and it will be a future challenge to make this precise.

Seen through the lens of general thermodynamics, this work shows that the concept of thermodynamic length can be extended to the nonequilibrium situation.
This is necessitated by the physically meaningful driving of the concentration variables $x$ instead of the extensive parameters $\eta$, which characterize the equilibrium system.
In regard to the original derivation in \cite{salamon1983}, where only a specific driving field on the space of $\eta$ parameters is considered, we have shown how the isometric embedding into a larger state space can yield a more detailed understanding of the driving kinetics and enable the calculation of explicit bounds.

The explicit analytical results obtained in this work shed new light on time-reversed driving:
Results 1 and 3 show that the optimal system trajectory for the driving from $x_1$ to $x_0$ will be the time-reversed trajectory caused by the optimal driving from $x_0$ to $x_1$.
However, the driving protocols will follows different paths (see Appendix \ref{app:crooks} for a discussion of time reversal illustrated with an example).
This leads to the conclusion that the approach taken in \cite{crooks2007} to compute the total dissipation for the driving between $x_0$ and $x_1$ is strictly valid only in the $T \rightarrow \infty$ limit for the trivial weight matrix $K=kI$ with a uniform rate constant $k \in \R_{>0}$.

In future work, it will be interesting to address the general case of an arbitrary driving protocol and without any assumptions on time scale separation or restrictions of the direction of the driving field.
Thereby, the results from \cite{sughiyama2021}, in particular the Pythagorean theorems, will be useful to evaluate the contribution to dissipation through chemical reactions.

However, there are more conceptual questions that have been raised by the presented analysis.
The weighted metric $g^K_X$ should be thought of as an Onsager matrix for the fluxes $j_D$ and the corresponding forces $\mu^e - \mu$.
This is in line with the interpretation of the metric as the generalized friction tensor given in \cite{crooks2012b}.
It will be a rewarding future challenge to work out the geometry of optimal driving for arbitrary Onsager matrices.
Moreover, the fact that the calculated optimal curve did not depend on the particular details of the driving kinetics and was of thermodynamical origin requires a thorough physical explanation.
It might be necessary to employ a mathematically more advanced framework to disentangle the contributions from kinetics and thermodynamics, which are mixed within the matrix $g^K_X$ in our current approach.

Apart from the theoretical insights and questions raised in this work, we have provided another tool to better understand biochemical reaction networks and to aid in the design and optimization of operational protocols in industrial chemical applications.

\begin{acknowledgments}
We thank Atsushi Kamimura and all other members of our lab for fruitful discussions. 
This research is supported by JSPS KAKENHI Grant Numbers 19H05799 and 21K21308, and by JST CREST JPMJCR2011 and JPMJCR1927.
\end{acknowledgments}

\appendix

\section{Geodesic equations} \label{app:geodesics}

On a Riemannian manifold $X$ with local coordinates $(x^1,\dotsc,x^n)$ and Riemannian metric $$g^{ij} = g\left(\frac{\partial}{\partial x^i},\frac{\partial}{\partial x^j}\right)$$ the geodesic equations read
\begin{align*}
    \frac{\dd^2 x^k}{\dt^2} + \sum_{i,j =1}^n \Gamma^k_{ij} \frac{\dd x^i}{\dt}\frac{\dd x^j}{\dt} = 0.
\end{align*}
Hereby, the Christoffel symbols are given by
\begin{align*}
    \Gamma^m_{ij} = \frac{1}{2} \sum_{k=1}^n \left[ \frac{\partial}{\partial x^i} g^{jk} + \frac{\partial}{\partial x^j} g^{ki} - \frac{\partial}{\partial x^k} g^{ij} \right] g_{km}
\end{align*}
and $g_{km}$ are the matrix elements of the inverse of $g = (g^{ij})$, which is given by a matrix in local coordinates.
See, for example, \cite{carmo1992} for more details.

\section{Optimal driving in linear CRN} \label{app:linearCRN}

A CRN is called linear if each reaction has exactly one reactant and one product.
Linear CRN with mass action kinetics are mathematically equivalent to Markov chain models on a graph, which are widely used in stochastic thermodynamics.
Whereas the latter strictly obey the conservation of probability, the conserved quantities in linear CRN can be driven.
In this appendix it is shown that the optimal driving on the steady state manifold of a linear quasi-thermostatic CRN, as described in Result \ref{res:3}, coincides with the optimal driving on $X$, given in Result~1.

Without loss of generality, consider a linear CRN with one linkage class \cite{feinberg2019}.
The stoichiometric matrix $S$ of the CRN is identical to the incidence matrix of the digraph with $n$ vertices given by the chemicals $X_1,\dotsc,X_n$ and directed edges between $X_i$ and $X_j$ iff there is a reaction $X_i \rightarrow X_j$.
Therefore $S^T$ has a one-dimensional kernel spanned by $u^* = (1~1~\dots~1)$ and there is one conserved quantity given by
\begin{align*}
    \eta = \sum_{i=1}^n x^i.
\end{align*}
The steady state manifold is given by $\Vss = \{x_{ss} \circ \exp(\eta^*) | \eta^* \in \R \}$, which yields the parametrization of all $x \in \Vss$ via the conserved quantity $\eta \in \R_{>0}$ as
\begin{align*}
    x = \frac{\eta}{\eta_{ss}} x_{ss},
\end{align*}
where $\eta_{ss} = \langle u^*, x_{ss} \rangle = \sum_{i=1}^n x^i_{ss}$.
Thus, the metric on $\D = \R_{>0}$ is given by \eqnref{eq:g_eta_explicit} as
\begin{align*}
    g_{\D}^K &= \left((\UU)^TD\UU\right)^{-1} (\UU)^T D K^{-1}\UU \left((\UU)^TD\UU\right)^{-1} \\
                &= \frac{1}{\eta \eta_{ss}} \sum_{i=1}^n x^i_{ss} k_i^{-1},
\end{align*}
which gives the Christoffel symbol $\Gamma = - \frac{1}{2\eta}$ and the geodesic equation for $\eta$
\begin{align*}
    \frac{\dd^2 \eta}{\dd t^2} -\frac{1}{2\eta} \left( \frac{\dd \eta}{\dd t} \right)^2 = 0.
\end{align*}
This equation has the explicit solution
\begin{multline} \label{eq:opt_driving_example}
    \eta(t) = - \left( \sqrt{\eta_1} - \sqrt{\eta_0} \right)^2 \frac{t}{T} \left( 1- \frac{t}{T} \right) + \\
     + \eta_1 \frac{t}{T} + \eta_0 \left( 1 - \frac{t}{T} \right),
\end{multline}
where $\eta_0 = \eta(0)$ and $\eta_1 = \eta(T)$.
The ideal driving on $\Vss$ between the points $x_0 = \frac{\eta_0}{\eta_{ss}} x_{ss}$ and $x_1 = \frac{\eta_1}{\eta_{ss}} x_{ss}$
thus follows the trajectory
\begin{align} \label{eq:curve_example}
    x(t) = \frac{\eta(t)}{\eta_{ss}} x_{ss}.
\end{align}
The total dissipation is given by
\begin{align*}
        \Sigma_{\Vss}^{min} &= T \sigma^{\min} = T \frac{\dd \eta(t)}{\dt} g_{\D}^K \frac{\dd \eta(t)}{\dt} \\
        &= \frac{4}{\eta_{ss} T} 
        \left( \sqrt{\eta_1} - \sqrt{\eta_0} \right)^2 \sum_{i=1}^n x^i_{ss} k_i^{-1} \\
        &= \frac{4}{T} \sum_{i=1}^n k_i^{-1} \left( \sqrt{\frac{\eta_1}{\eta_{ss}} x^i_{ss}} - \sqrt{\frac{\eta_0}{\eta_{ss}} x^i_{ss}} \right)^2 \\
        &= \frac{4}{T} \sum_{i=1}^n k_i^{-1} \left( \sqrt{x^i_1} - \sqrt{x^i_0} \right)^2.
\end{align*}
This dissipation $\Sigma_{\Vss}^{min}$ is equal to the lower bound $\Sigma_X^{min}$ stated in Result \ref{res:2} and thus the driving on the equilibrium manifold coincides with the optimal driving for the fast driving regime given by \eqnref{eq:opt_driving}.
This can also be verified by directly substituting \eqnref{eq:opt_driving_example} into \eqnref{eq:curve_example}.

\section{Reversal of driving} \label{app:crooks}

\begin{figure}[ht]
\includegraphics[width=0.5\textwidth, clip]{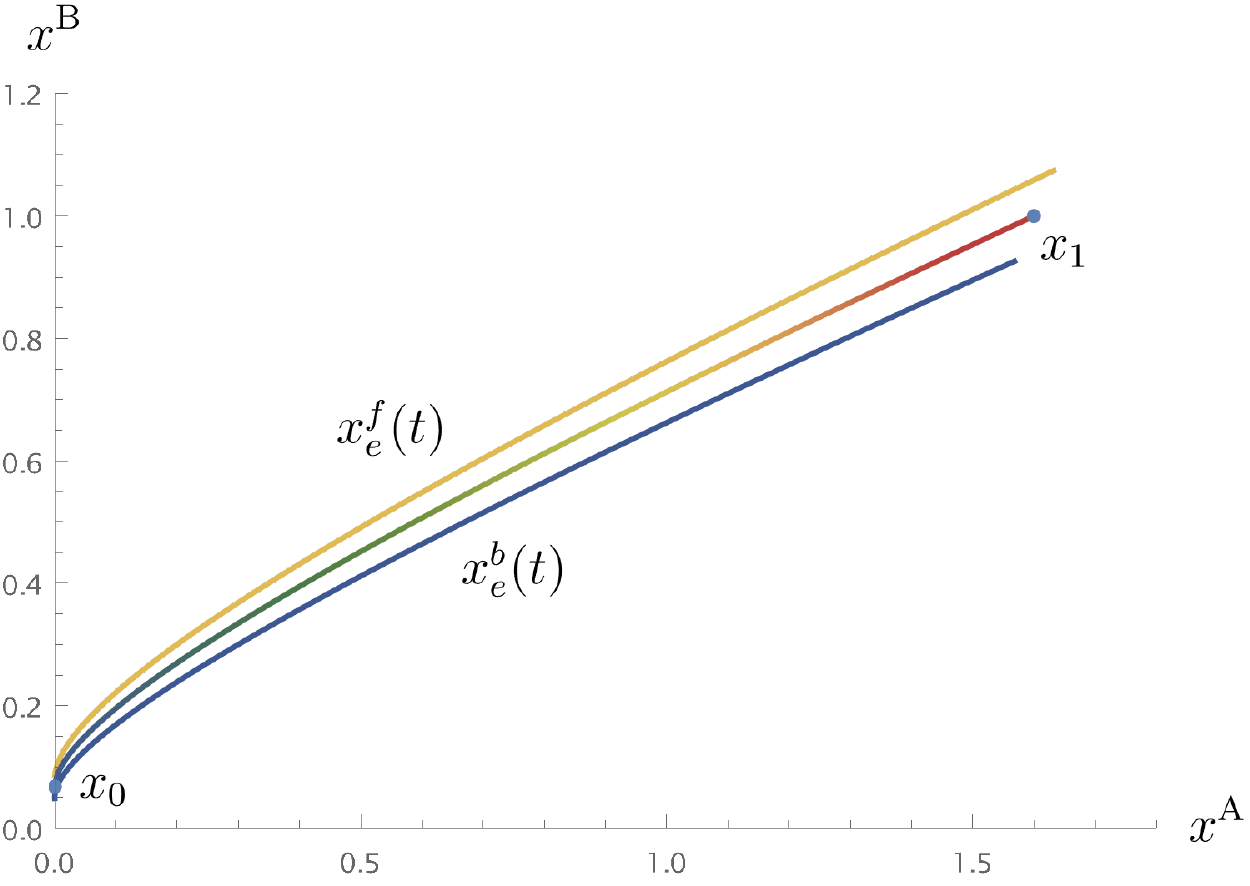}
\caption{Reversal of driving: an example.
    The example from Fig. \ref{fig:example} with the assumption that no reactions take place between A and B is used here.
    The optimal protocol $x^f_e(t)$ for the driving from $x_0$ to $x_1$ is shown in yellow and the optimal protocol $x^b_e(t)$ for the driving from $x_1$ to $x_0$ is shown in blue.
    }
\label{fig:crooks}
\end{figure}

The Results 1 and 3 imply that a system which is optimally driven from $x_0$ to $x_1$ will follow the same trajectory as the optimally driven system from $x_1$ to $x_0$ with time-reversal:
Denoting the former trajectory by $x^f(t)$ and the latter by $x^b(t)$, the relation $x^b(t) = x^f(T-t)$ for all $t \in [0,T]$ holds.
However, the same results show that this is not true for the driving protocols $x_e^f(t)$ and $x_e^b(t)$.
They lie on different curves and cannot be identified under time reversal, even if one would introduce a lag time.
This is illustrated in Fig. \ref{fig:crooks}.
Therein, the example from Sec. \ref{sec:toric} is considered with the additional assumption that there are no chemical reactions between $A$ and $B$ and thus that all of $X$ is the equilibrium manifold. 
The driving protocols $x_e^f(t)$ and $x_e^b(t)$ are different non-intersecting curves.

For $x_e^f(t)$ and $x_e^b(t)$ to become the same curves, it is necessary and sufficient that the matrix $K$ is proportional to the identity matrix and that the time limit $T \rightarrow \infty$ is taken.
In this case, they are identified with the geodesic $x(t)$.

\clearpage

\bibliography{apssamp}

\end{document}